\documentclass[prb,aps,twocolumn,showpacs,superscriptaddress,raggedbottom]{revtex4}

\usepackage{graphicx,bm}
\usepackage{dcolumn}
\usepackage{amsmath}

\def\i#1{\textit{#1}}
\def\v#1{\textbf{#1}}
\def\b#1{\mathbf{#1}}
\def\mb#1{\mbox{\boldmath$#1$}}
\def\nn{\nonumber \\}

\def\eq#1{Eq.~(\ref{#1})}
\def\fig#1{Fig.~\ref{#1}}

\def\be{\begin{equation}}
\def\ee{\end{equation}}
\def\bea{\begin{eqnarray}}
\def\eea{\end{eqnarray}}

\begin{document}

\title{Generation of spin current and polarization under dynamic gate control
of spin-orbit interaction in low-dimensional semiconductor systems}

\author{C.~S. Tang}
\affiliation{Physics Division, National Center for Theoretical
Sciences, P.O. Box 2-131, Hsinchu 30013, Taiwan}
\author{A.~G. Mal'shukov}
\affiliation{Institute of Spectroscopy, Russian Academy of Science,
142190, Troitsk, Moscow oblast, Russia}
\author{K.~A. Chao}
\affiliation{Solid State Theory Division, Department of Physics,
Lund University, S-22362 Lund, Sweden}

\begin{abstract}
Based on the Keldysh formalism, the Boltzmann kinetic equation and
the drift diffusion equation have been derived for studying spin
polarization flow and spin accumulation under effect of the time
dependent Rashba spin-orbit interaction in a semiconductor quantum
well. The time dependent Rashba interaction is provided by time
dependent electric gates of appropriate shapes. Several examples of
spin manipulation by gates have been considered. Mechanisms and
conditions for obtaining the stationary spin density and the induced
rectified DC spin current are studied.
\end{abstract}





\maketitle

\section{introduction}

Spin transport in semiconductor heterostructures has recently
attracted much attention owing to a perspective of its practical
application for quantum computing and
communications.~\cite{Loss02,Wolf01,Rev.Mod.Phys} On the way of
fundamental studies and applications of the spin transport the key
problem is how to generate, detect, and manipulate the electron
spin polarization. Many ideas have been proposed to achieve
control of the spin using magnetic materials, external magnetic
fields and optical excitation (for a review see
Ref.~\onlinecite{Rev.Mod.Phys}). At the same time, a challenging
goal retains to employ only the electric control integrated by
means of gates into a high mobility semiconductor heterostructure.

A promising opportunity on this way opens due to the strong Rashba
spin-orbit interaction (SOI) in narrow gap two-dimensional (2D)
electron systems.~\cite{Rash84} An important feature of this
interaction is its tunability which can be achieved by varying the
gate voltage.~\cite{Nitta97,Grun00} The field effect transistor was
the first proposal utilizing this phenomenon.~\cite{Das} The gate
control of the spin current employing the Aronov-Casher
effect~\cite{Mathur} was considered in Ref.~\onlinecite{Shlyapin}.
The electric dipole spin resonance controlled by the time dependent
gate was studied in Ref.~\onlinecite{RashbaLev}.

The SO interaction effect alone, or in combination with the external
electric field, allows to approach an important goal to create spin
currents and spin polarization by entirely electrical means, not
involving any of the magnetic materials or optical excitation. One
of the examples is the recently predicted~\cite{Sinova03} (see also
Ref.~\onlinecite{Hirsch}) and observed~\cite{Wunderlich,Kato}
spin-Hall effect in 2D and 3D electron and hole gases, where the
spin current is driven by the electric field. The closely related
phenomenon is the spin polarization of 2DEG in response to the
parallel electric field.~\cite{Edelstein} Another method utilizes
the time dependent gate to modulate the shape of a quantum
dot~\cite{Sharma03} or the strength of the SO coupling constant in
1D~\cite{Gover02,Malsh03} and 2D systems.~\cite{Malsh03}  In the
latter case an efficient spin current generation can be attained in
the presence of high frequency (hundreds MHz or higher) gate-voltage
variations. The physics of this phenomenon is simple. The Rashba
spin-orbit interaction resembles an interaction with a spin
dependent gauge field. Therefore, its time derivative produces a
motive force on electrons, similar to the electromotive force from
the time dependent electric vector potential. The important
difference is, however, that the force created by SOI acts in
opposite directions for oppositely polarized spins. Such a method of
spin current generation is convenient for implementation in
conventional semiconductor heterostructures. By a proper choice of
the gate shape it allows to create and rectify the AC spin current
or accumulate the spin polarization at a given location. However,
the theory in Ref.~\onlinecite{Malsh03} is restricted to the
spatially homogeneous case. Therefore, within this theory one can
not properly consider boundary effects, as well as the spin current
generation due to a time dependent gate of the small area. Also, it
is impossible to study any effects of the spin current generation
accompanied by spin accumulation.

In the present work we develop a theory which is based on the
Boltzmann transport equations. This approach is quite universal
and it allows us to study the spin transport under an arbitrary
space-time dependent SO interaction, providing that a
characteristic scale of this dependence is within the range of
applicability of the semiclassical approximation. This means that
the spatial variations of the SO interaction are assumed to have
the scale $\Delta r$ larger than the electron Fermi wavelength and
the scale of its time variations $\Delta t \gg \hbar/E_F$, where
$E_F$ is the Fermi energy. For even more larger time-space scales,
such as $\Delta r \gg l$ and $\Delta t \gg \tau$, where $l$ and
$\tau$ are, respectively, the mean free path and the mean free
scattering time, we use the Boltzmann equation to derive the
drift-diffusion equation. Within this theory we consider the
following problems:
\begin{enumerate}
    \item Spin current generation by a finite size time dependent gate.
    In this case the spin polarization is pumped by the gate into
    2D regions adjacent to it. This polarization further diffuses
    away from the gate, as well as in the backward direction.
    \item  Spin current generation in a gas confined within an infinite
    2D strip. In such a
    geometry the generated spin current with the polarization perpendicular
    to  boundaries flows freely along the strip. At the same
    time, spins polarized parallel to the boundaries are
    accumulated near them, opposite spins near opposite banks.
    \item Rectification of the AC spin current. It will be shown
    that the spin polarization generated by a finite size AC gate in some cases
    contains the DC component due to periodic variations of the electron density under
    the gate. The DC component appears as a result of interplay of two
periodic processes:
    oscillations of the SO coupling constant and oscillations of the 2D electron
density.

\end{enumerate}

This paper is organized as follows. In Sec. II, the Boltzmann
equation is derived for calculation of the spin transport in the
presence of the time-dependent SO interaction. In Sec. III, we apply
the Boltzmann equation to some of the spin transport problems. The
derivation of the drift-diffusion equation and its applications are
presented in Sec. IV. The results of this work are summarized in
Sec. V.


\section{Boltzmann transport theory}
\label{BTT}

We consider spin transport in a noninteracting 2DEG confined
within a narrow-gap semiconductor quantum well (QW). The 2DEG is
applied atop by a gate under time dependent bias. The inversion
asymmetry of the quantum well confining potential is controllable
by an external gate, and hence electrons experience a tunable
Rashba-type SO interaction.~\cite{Nitta97, Grun00} The system is
therefore described by the effective mass Hamiltonian
\begin{eqnarray}\label{H}
H(t) &=& \frac{\b{p}^2}{2m^*} + \b{M}\cdot \frac{1}{2}\left[
\alpha(\b{r},t)\b{p} + \b{p}\alpha(\b{r},t) \right] \nn && +
U(\b{r},t),
\end{eqnarray}
where $m^*$ represents the effective mass, $\b{p} = (p_x,p_y)$ is
the electron momentum in the 2DEG plane, and $\alpha(\b{r},t)$
denotes the time-dependent Rashba coupling constant.
$\alpha(\b{r},t)$ contains both a static part $\alpha_0$ and a
dynamic part $\alpha_1(\b{r},t)$ due to the time dependent gate.
In addition to time dependence, $\alpha_1$ can also vary in space
depending on the gate configuration and applied bias. We denote
$\b{M}=\mb{z}\times\mbox{\boldmath$\sigma$}$ where $\mb{z}$ is the
unit vector along the growth direction and
$\mbox{\boldmath$\sigma$}=\{\sigma_x,\sigma_y,\sigma_z \}$ is the
vector of the Pauli matrices. Further, $U(\b{r},t)$ indicates the
external potential energy of an electron in the electric field
produced by the gate. The static part of the Rashba coupling
constant lifts the spin degeneracy resulting in a
momentum-dependent spin-splitting of the conduction band, namely
\begin{equation}
\varepsilon_{\pm}(\b{k}) = \frac{\b{k}^2}{2m^*} \pm \alpha_0 k .
\end{equation}
For 2DEG density of interest the Rashba spin-splitting energy
$\Delta_0 = 2 \alpha_0 k$ will be assumed to be small compared to
the Fermi energy, $\Delta_0 \ll E_F$. In the following, we set
$\hbar = 1$ for convenience.

Let us consider the time-modulated nonequilibrium electron system
at temperature regime $\mathrm{T} \ll E_F$, assuming the
semiclassical conditions to be satisfied.  Hence, the time
intervals $\Delta t$ and distances $\Delta r$ over which all
quantities vary significantly satisfy the inequalities $E_F\Delta
t  \gg 1$ and $p_F\Delta r \gg 1$.  These conditions are necessary
for derivation of the kinetic equation. We start this derivation
from introducing the matrix of nonequilibrium Green's function in
the Keldysh space:~\cite{Keldysh64}
\begin{equation}\label{hatG}
\hat{\mathcal{G}}_{\alpha\beta}(1,2) = \left[
{\begin{array}{*{20}c}
   {G_{\alpha\beta}^{--}(1,2)} & {G_{\alpha\beta}^{-+}(1,2)}  \\
   {G_{\alpha\beta}^{+-}(1,2)} & {G_{\alpha\beta}^{++}(1,2)}  \\
\end{array}} \right] \, .
\end{equation}
where $1$ and $2$ denote two points $\b{r}_1,t_1$ and $\b{r}_2,t_2$,
in the time-space and $\alpha, \beta$ are spinor indices. We will
assume that besides the time dependent forces in the Hamiltonian
(\ref{H}), the electrons are also subject to scattering from
randomly distributed static impurities. This scattering can be taken
into account by introducing the self energy
\begin{equation}\label{hatSigma}
\hat{\mathit{\Sigma}}_{\alpha\beta} = \left[
{\begin{array}{*{20}c}
   {\Sigma_{\alpha \beta }^{--} } & {\Sigma_{\alpha \beta }^{-+} }  \\
   {\Sigma_{\alpha \beta }^{+-} } & {\Sigma_{\alpha\beta}^{++} }  \\
\end{array}} \right]\, .
\end{equation}
The corresponding matrix Dyson equation is of the form
\begin{eqnarray} \label{Dyson1}
\hat{\mathcal{G}}_{\alpha\beta}(1,2) &=&
\hat{\mathcal{G}}_{\alpha\beta}^0(1,2) \nn &+& \int d4 d3
\left[\hat{\mathcal{G}}_{\alpha\gamma}^0(1,4)
\hat{\mathit{\Sigma}}_{\gamma\delta}(4,3)
\hat{\mathcal{G}}_{\delta\beta}(3,2) \right]
\end{eqnarray}
or the conjugate form
\begin{eqnarray}\label{Dyson2}
\hat{\mathcal{G}}_{\alpha\beta}(1,2) &=&
\hat{\mathcal{G}}_{\alpha\beta}^0(1,2) \nn &+& \int d3 d4
\left[\hat{\mathcal{G}}_{\alpha\gamma}(1,3)
\hat{\mathit{\Sigma}}_{\gamma\delta}(3,4)
\hat{\mathcal{G}}_{\delta\beta}^0(4,2) \right],
\end{eqnarray}
where the functions in the integrand are matrices both in real
space and in spin space being combined by the rule of matrix
multiplication. Acting on the Dyson equation from the left (or
from the right on its conjugate) by the operator
\begin{equation}
\left[G_{\alpha\beta}^0(j)\right]^{-1} \equiv
i\frac{\partial}{\partial t_j}\delta_{\alpha\beta} -
H_{\alpha\beta}(t_j)
\end{equation}
where the suffix $j$ = $1$ or $2$ indicates the differentiation is
with respect to the variables $t_j$ and $\mb{r}_j$. After some
algebra, we obtain the equation~\cite{Lifshitz}
\begin{widetext}
\begin{eqnarray}\label{kin12}
\left( \hat{\mathcal{I}}_{\text{sc}}\right)_{\alpha \beta } &=&
-i\left( {\frac{\partial } {\partial t_1} + \frac{\partial }
{\partial t_2}} \right)\hat{\mathcal{G}}_{\alpha \beta }(1,2)  +
\frac{1} {{2m}}\left( {\Delta _2  - \Delta _1 }
\right)\hat{\mathcal{G}}_{\alpha \beta }(1,2) + \left[ U(1) -U(2)
\right] \hat{\mathcal{G}}_{\alpha \beta }(1,2) \nn && - i\left[
{\alpha(2) \mb{\nabla}_2  \cdot \left( {\hat{\mathcal{G}}_{\alpha
\gamma}(1,2)\b{M}_{\gamma\beta}} \right) + \alpha(1) \mb{\nabla} _1
\cdot \left( {\b{M}_{\alpha\gamma}\hat{\mathcal{G}}_{\gamma \beta
}(1,2)} \right)} \right] \nn &&- \frac{i} {2}\left[ {\mb{\nabla} _2
\alpha(2) \cdot \left( {\hat{\mathcal{G}}_{\alpha \gamma
}\b{M}_{\gamma\beta}} \right) + \mb{\nabla} _1 \alpha(1) \cdot
\left( \b{M}_{\alpha\gamma}\hat{\mathcal{G}}_{\gamma\beta }(1,2)
\right)} \right],
\end{eqnarray}
\end{widetext}
which is equivalent to the set of integro-differential equations
for the component Green functions.  Here the scattering integral
is defined by~\cite{Lifshitz}
\begin{eqnarray}\label{I12}
\left({\hat{\mathcal{I}}}_{\text{sc}} \right)_{\alpha\beta}
 &=&  \int d3\,
\left[\hat{\mathit{\tau}}
\hat{\mathit{\Sigma}}_{\alpha\gamma}(1,3)\hat{\mathcal{G}}_{\gamma\beta}(3,2)
\right. \nn && \left. - \hat{\mathcal{G}}_{\alpha\gamma}(1,3)
\hat{\mathit{\Sigma}}_{\gamma\beta}(3,2)\hat{\mathit{\tau}} \right]
,
\end{eqnarray}
where $\hat{\mathit{\tau}}$ is the matrix, such that
$\tau^{++}=-\tau^{--}=1$ and $\tau^{+-}=\tau^{-+}=0$.

In the semiclassical regime, it is convenient to transform the
variables to the Wigner coordinates: $\b{r} = \b{r}_1 - \b{r}_2$,
$\b{R} = \frac{1}{2}(\b{r}_1 + \b{r}_2)$; $t = t_1 - t_2$, $T=
\frac{1}{2}(t_1 + t_2)$. The difference variables $\b{r},t$ vary
on a microscopic scale, while the center variables $\b{R}$ and $T$
are macroscopic variables. In the quasiclassic approximation the
Green functions and self energies vary slowly with respect to the
center variables. Accordingly, only linear gradient expansions
will be taken into account in (\ref{kin12}).~\cite{Haug96}It is
appropriate to define the space-time Fourier transform to the fast
variables
\begin{eqnarray}\label{Fourier}
G^{ij}_{\alpha\beta}(\b{R},T;\b{r},t) &=& \sum_{\b{k}} \int \frac{d\omega}{2\pi}
G^{ij}_{\alpha\beta}(\b{R},T;\b{k},\omega)\nn && \times \exp
\left[ i \left( \b{k}\cdot \b{r} - \omega t \right) \right],
\end{eqnarray}
where the uppercase indices $i,j$ take the values ``$+$" or ``$-$".
Denoting as $G^{ij}(\b{R},T;\b{k},\omega)$ a matrix with elements
$G^{ij}_{\alpha\beta}(\b{R},T;\b{k},\omega)$  and making the
gradient expansion of (\ref{kin12}) one can write the quantum
kinetic equation in the form
\begin{widetext}
\begin{eqnarray}\label{qke1}
iI^{ij}_{\text{sc}} &=& \frac{\partial G^{ij}}{\partial T} + \b{v}
 \cdot \mb{\nabla}_{\bf R} G^{ij} + i
\alpha \left[ \b{k} \cdot \b{M}, G^{ij} \right] +
\frac{1}{2}\alpha \left\{ \b{M},\mb{\nabla}_{\bf R} G^{ij}
\right\}  \nn && - \frac{1}{2} \mb{\nabla}_{\bf R}\alpha \cdot
\left\{ \b{k} \cdot \b{M}, \mb{\nabla}_{\bf k} G^{ij} \right\} +
\frac{1}{2} \frac{\partial \alpha}{\partial T}\left\{ \b{k} \cdot
\b{M},\frac{\partial G^{ij}}{\partial \omega} \right\}-
{\mb{\nabla}}U\cdot \mb{\nabla}_{\bf k} G^{ij} + \frac{\partial
U}{\partial T}\frac{\partial G^{ij}}{\partial \omega} ,
\end{eqnarray}
\end{widetext}
where $\b{v} = \b{k}/m^*$ and $\{A,B\}$ denotes the anticommutator.
The scattering part of this equation is obtained from (\ref{I12}) in
the leading order of the gradient expansion with respect to the
``slow" $\b{R},T$ variables.~\cite{Lifshitz} In calculations below
we will employ the three Green functions. There are $G^{-+}$,
retarded $G^{r}$ and advanced $G^{a}$ functions. $G^{r}$ and $G^{a}$
are defined by the equations
\begin{eqnarray}\label{retarded}
G^{r} &=& G^{--}-G^{-+}= G^{+-} - G^{++}\nn G^{a} &=&
G^{-+}-G^{++}= G^{--}-G^{+-}.
\end{eqnarray}
The same equations are valid also for the retarded and advanced
self energies with the only difference that $\Sigma^{+-}$ and
$\Sigma^{-+}$ enter with the signs opposite to signs of $G^{+-}$ and
$G^{-+}$ in (\ref{retarded}). In terms of these functions the
corresponding scattering parts of (\ref{qke1}) are written as
\begin{equation}\label{I-+}
I^{-+}_{\text{sc}}=-\Sigma^{r} G^{-+} + G^{-+}\Sigma^{a} -
G^{r}\Sigma^{-+} + \Sigma^{-+} G^{a},
\end{equation}
\begin{equation}\label{Ira}
I^{r}_{\text{sc}}=-\left[\Sigma^{r},G^{r}\right] ;I^{r}_{\text{sc}}=
-\left[\Sigma^{a},G^{a}\right]\, .
\end{equation}

In order to determine the self energy we assume, for simplicity,
that the impurity scattering potential is isotropic, spin
independent and short range. Hence, it can be simply written as
$V(\b{r})=V\delta(\b{r})$ so that the corresponding Born scattering
amplitude does not depend on the electron wavevector. In  the
quasiclassic approximation ignoring weak localization effects the
averaged over random impurity positions self energy is thus given
by~\cite{Rammer}
\begin{equation}\label{Sigma}
\Sigma^{ij}=(-1)^{i+j}|V|^2 \sum_{\b{k}}G^{ij}(\b{R},T;\b{k},\omega).
\end{equation}
One can see that the so defined self energy does not depend on the
wavevector. In the thermodynamic equilibrium the unperturbed
retarded and advanced Green functions are easily found from
(\ref{Sigma}) and (\ref{H}):
\begin{equation}\label{Gr}
G^{r}_0 = G^{a\dag}_0 = \left(\omega - H_0 + i\Gamma \right)^{-1},
\end{equation}
where $H_0= (k^2/2m^*) + \alpha_0 \b{M}\cdot \b{k}$. The elastic
scattering rate $\Gamma$ is obtained from (\ref{Sigma}).
Substituting (\ref{Gr}) into (\ref{Sigma}) one easily finds
$\Gamma=\pi |V|^2 N_F$, where $ N_F=m^*/(2\pi)$ is the one particle
2D density of states.~\cite{Rammer} The ``$-+$" unperturbed Green
function can be found from the equation~\cite{Lifshitz}
\begin{equation}\label{G-+0}
G^{-+}_0=-n_{F}(\omega)(G^{r}_0 - G^{a}_0),
\end{equation}
where $n_{F}(\omega)$ is the equilibrium Fermi distribution.

The $2\times 2$ matrix  Wigner distribution function is defined as
$-iG^{-+}(\b{R},T;\b{k},\omega)$. At the same time, the density
matrix which is used for calculations of observable physical
quantities, for example, spin density and spin current, can be
obtained from the Wigner function by integration over $\omega$.
Usually, the Boltzmann equation for this function is obtained by
integration over $\omega$ of the quantum kinetic equation, such as
the ``$-+$" component of \eq{qke1}. In our case, however, this
method can not be used because of the term proportional to $\partial
\alpha/\partial T$, which vanishes after integration. Moreover, it
becomes a not simple task to write the scattering part in terms of
the density function. Similar problem arises when one derives the
Boltzmann equation for the system driven out of equilibrium by the
electric field represented by a time dependent vector potential.
This difficulty is resolved by shifting variables from wavectors to
kinematic momenta or to velocities.~\cite{Rammer} In the problem
considered here such a shift is not much helpful, because the
velocity $\b{k}/m^* + \alpha(\b{R},T) \b{M}$ is a matrix in the spin
space, not a number. We will employ a different method. Let us
represent the Wigner function in the form
\begin{equation}\label{anzatz}
-iG^{-+}=i n_{F}(\omega)(G^{r} - G^{a}) + F.
\end{equation}
At $U(\b{R},T)$=0 and $\alpha(\b{R},T)=\alpha_0$ the first term in
the right-hand side turns into the unperturbed Wigner function.
Hence, the $F$-function is not zero only due to deviation of the
system from the original homogeneous thermodynamically equilibrium
state. Assuming this deviation to be small, we will linearize
\eq{qke1}, omitting all products of $F$ with terms containing the
time and space derivatives of $\alpha$ and $U$. After substitution
of \eq{anzatz} to the ``$-+$" component of \eq{qke1}, taking into
account \eq{I-+} and  the corresponding equations for $G^r$ and
$G^a$ with the scattering terms given by \eq{Ira}, we arrive to the
following equation for $F$:
\begin{widetext}
\begin{eqnarray}\label{qkeF}
\mathrm{St}[F]&=& \frac{\partial F}{\partial T} + \b{v}
 \cdot \mb{\nabla}_{\bf R} F + i
\alpha \left[ \b{k} \cdot \b{M}, F \right] + \frac{1}{2}\alpha
\left\{ \b{M},\mb{\nabla}_{\bf R} F \right\}  \nn && +\frac{i}{2}
\frac{\partial \alpha}{\partial T}\left\{ \b{k} \cdot \b{M},G^r -
G^a \right\} \frac{\partial n(\omega)}{\partial \omega} +i
\frac{\partial U}{\partial T}(G^r - G^a) \frac{\partial
n(\omega)}{\partial \omega},
\end{eqnarray}
\end{widetext}
where $\mathrm{St}[F]$ is given by
\begin{equation}\label{IF}
\mathrm{St}[F]= \Sigma^{r}F - F\Sigma^{a} - G^{r}\Sigma(F) +
\Sigma(F) G^{a},
\end{equation}
with $\Sigma(F)$ defined as $\Sigma(F)=|V|^2 \sum_{\b{k}}F$.

A good approximation to the functions $G^{r}$ and $G^{a}$ are the
local equilibrium functions $G^{r}_l$ and $G^{a}_l$
defined by
\begin{equation}\label{Gl}
G^{r}_l=G^{a\dag}_l=(\omega - H_l + i\Gamma)^{-1},
\end{equation}
where $H_l= (k^2/2m^*) + \alpha(\b{R},T) \b{M}\cdot \b{k} +
U(\b{R},T)-E_F$. The local functions has the same form as the
equilibrium functions (\ref{Gr}) with the electron energy and spin
splitting parametrically dependent on time an space. It can be seen
from (\ref{qke1}), using (\ref{retarded}), (\ref{Ira}) and
(\ref{Sigma}) that the corrections to the local functions are
proportional to the Green function gradients times the small
parameter $\alpha/v_F$, where $v_F$ is the Fermi velocity. Such a
small parameter can be important for the spin-Hall
effect~\cite{Sinova03}, but not for the considered here case of the
spin current driven by the time dependent SO coupling constant. One
can further simplify Eq. (\ref{qkeF}) omitting proportional to this
parameter terms in (\ref{qkeF}), such as $ \alpha \left\{
\b{M},\mb{\nabla}_{\bf R} F \right\}$, which enters together with
the much bigger $\b{v} \cdot \mb{\nabla}_{\bf R} F$. After this
simplification the linear operator acting upon $F$ in the right-hand
side of Eq.~(\ref{qkeF}) decouples into scalar and spin dependent
parts. Further, since we are interested in the spin transport driven
by the term in (\ref{qkeF}) proportional to the time derivative of
$\alpha$, one can ignore the term containing $\partial U/ \partial
T$, because it contributes additively into the linear response and
gives rise to the spin-Hall effect which will not be considered in
the present work. It should be noted that the linearization is
undertaken only with respect to the time derivatives, while local
values of $\alpha$ and $U$ entering into $G^{r(a)}_l$ can vary
noticeably within macroscopic time-space scales.

Now let us take a look at the scattering part (\ref{IF}). One can
easily see that since SOI is an odd function with respect to
$\b{k}$, the self-energies $\Sigma^{r,a}$ calculated from
(\ref{Sigma}) with the local equilibrium functions (\ref{Gl}) are
contributed only by the scalar parts of these functions.
Therefore, omitting the small corrections of the order of
$\Delta/E_F$, where $E_F$ is the Fermi energy, from (\ref{Gl}),
(\ref{Sigma}) and (\ref{retarded}) we get
\begin{equation}\label{Itau}
\Sigma^{r}F - F\Sigma^{a} = -2i \Gamma F.
\end{equation}
The other terms of (\ref{IF}), the ones which contain $\Sigma(F)$,
can be also simplified. We notice that the spin dependent parts of
$G^{r(a)}_l$ contribute effectively with the small parameter
$\Delta/E_F$. Although small, these terms provide coupling of the
particle density represented by the scalar $Tr[\Sigma(F)]$ to the
spin distribution function associated with $Tr[\bm{\sigma}F]$ and
vice versa. They are important in the spin-Hall
effect~\cite{Mischenko,Burkov}. In our case we can ignore them.
Taking into account all these simplifications we are ready to derive
the Boltzmann equation for the particle distribution function in the
space of particle coordinates and momenta. This function is obtained
by integration over $\omega$ of Eq.~(\ref{anzatz})~\cite{Lifshitz}
with $G^{r,a}$ given by (\ref{Gl}). In the leading approximation we
thus get
\begin{eqnarray}\label{f1}
-i\int \limits_{-\infty }^\infty \frac{d\omega}{2\pi} G^{-+} &=&
n_{Fl} + \alpha \b{M} \cdot \b{k} \frac{\partial n_{Fl}}{\partial
E} \nn && + f_{\b{k}}(\b{R},T),
\end{eqnarray}
where $n_{Fl}=n_F\left( E+U(\b{R},T)\right)$ is the local Fermi
distribution function, where the coordinate and time dependent
potential energy is added to the electron kinetic energy $E$. The
function $f_{\b{k}}(\b{R},T)$ is defined by
\begin{equation}\label{f}
f_{\b{k}}(\b{R},T) = \int\limits_{-\infty }^\infty
\frac{d\omega}{2\pi} F(\b{R},T;\b{k},\omega).
\end{equation}
After integration Eq.~(\ref{qkeF}) over frequency we get the
Boltzmann equation in the form
\begin{eqnarray}\label{qkef}
\mathrm{St}[f]&=& \frac{\partial f_{\b{k}}}{\partial T} + \b{v}
 \cdot \mb{\nabla}_{\bf R} f_{\b{k}} + i
\alpha(\b{R},T) \left[ \b{k} \cdot \b{M}, f_{\b{k}} \right] \nn &&
+\b{k} \cdot \b{M} \frac{\partial \alpha(\b{R},T)}{\partial
T}\frac{\partial n_{Fl}}{\partial E} ,
\end{eqnarray}
The scattering term is obtained as
\begin{equation}\label{St}
\mathrm{St}[f]=\frac{2\pi}{\tau m^*}
\sum_{\b{k}'}\delta(E^{\prime}-E)f_{\b{k}'} -
\frac{f_{\b{k}}}{\tau},
\end{equation}
where $\tau=1/2\Gamma$ is the elastic scattering time.

\section{simple examples of spin transport under the time dependent gate}

In this section, we will employ \eq{qkef} to investigate electron
transport properties of two-dimensional electron systems in the
presence of time-dependent SOI. Two simple examples of application
of the Boltzmann equation will be considered: spin transport driven
by a homogeneous infinite gate and the ballistic transport due to a
narrow time dependent gate.

Before to proceed with these examples, let us define the spin
current and spin density in terms of $f_{\b{k}}$. According to
definition of $G^{-+}$,~\cite{Lifshitz} the spin distribution in the
space of particle coordinates and momenta is given by the spin
dependent part of (\ref{f1}). Therefore, the spin density is
obtained by integration of nonscalar terms in \eq{f1} over $\b{k}$.
Taking into account that the second term in \eq{f1} turns to zero
after averaging over $\b{k}$ directions, we get the spin density
\begin{equation}\label{P}
P^i(\b{R},T)=\frac{1}{2}\sum_{\b{k}}Tr[\sigma^i
f_{\b{k}}(\b{R},T)].
\end{equation}
The spin current definition is based on the one-particle spin-flux
operator $\frac{1}{4} \{ \sigma^i, \mathrm{v} \}$, where the
velocity operator $\mathrm{v}=\b{v} + \alpha \b{M}$. Hence, using
Eq.~(\ref{f1}), the spin current density can be written in the
form
\begin{eqnarray}\label{Is}
\b{{\cal J}}^i(\b{R},T) &=& \frac{1}{4} \sum_{\b{k}}Tr \left[\{
\mathrm{v},\sigma^i\}(n_{Fl} + \alpha \b{M} \cdot \b{k}
\frac{\partial n_{Fl}}{\partial E})\right] \nn && + \frac{1}{4}
\sum_{\b{k}} Tr\left[ \left\{
\mathrm{v},\sigma^i\right\}f_{\b{k}}(\b{R},T)\right].
\end{eqnarray}
It is easy to see by a direct calculation that the first sum is
zero. Therefore, only the nonequilibrium part of the spin current
associated with $f_{\b{k}}(\b{R},T)$ contributes to (\ref{Is}). In
this connection, it should be noted that the equilibrium spin
current was found to be nonzero in a homogeneous 2D
gas.~\cite{Rashbascurr}  This is due to quantum effects which are
beyond the semiclassical approach used in the present work.

\subsection{Homogeneous case}

As a simple, yet nontrivial application, we consider the case of a
large AC-biased gate such that the time-dependent region can be
treated homogeneously. In this case one can omit $ \mb{\nabla}_{\bf
R}f_{\b{k}}$ in \eq{qkef} and look for a solution of \eq{qkef} in
the form $f_{\b{k}}=A(k) \b{M} \cdot \b{k}$. Since
$f_{\b{k}}=-f_{-\b{k}}$, the first term in \eq{St} turns to zero and
the solution of (\ref{qkef}) is easily obtained as
\begin{equation}\label{homo}
f_{\b{k}}= \b{M} \cdot \b{k}\frac{i\Omega
\b{\tau}}{1-i\Omega\b{\tau}} \alpha(\Omega)\frac{\partial
n_{Fl}}{\partial E},
\end{equation}
where $\alpha(\Omega)$ is the Fourier transform of $\alpha(T)$ at
the frequency $\Omega$. For simplicity we have assumed that the
electron density, as well as $\partial n_{Fl}/\partial E$, do not
change in time. On the other hand this dependence can be important
for obtaining the rectified DC spin current. This effect will be
discussed in the next section. Substituting (\ref{homo}) to
(\ref{Is}) we find the spin current expression in accordance with
Ref.~\onlinecite{Malsh03}
\begin{equation}\label{sc}
{\cal J}_j^i (\Omega) = \frac{1}{2}\frac{\rho\Omega}{ \Omega +
2i\Gamma }\varepsilon^{ij3}\alpha(\Omega),
\end{equation}
where $\rho =2 \sum_{\bf k} n_F(k)$ is the electron density, and the
indices $i$ and $j$ denote the spin polarization and direction of
the current flow, respectively. With the spin distribution function
(\ref{homo}) the spin polarization obtained from (\ref{P}) is
$\mb{P}=0$. Hence, in the homogeneous case no spin polarization is
induced.

\subsection{Current generation by a narrow gate}

We now consider the case of a narrow AC-biased gate, which is
infinitely long in $\mb{y}$ direction while the width $w$ in
$\mb{x}$ direction is much smaller than the electron mean free path
$l$, so that, except a small number of particles moving in the
$\mb{y}$ direction the motion of electrons under the gate is
ballistic. We are interested in the spin current flowing in the
$\mb{x}$ direction. Since $M_x=-\sigma_y$, it follows from
(\ref{qkef}) that this current is polarized along the $\mb{y}$-axis.
Neglecting the scattering term we obtain from Eqs. (\ref{qkef}) and
(\ref{Is})
\begin{eqnarray}\label{narrow}
{\cal J}_x^y\left(q_x,\Omega\right)  &=& \frac{m^{*2}}{4\pi}
 \int dE \left[\frac{\partial\alpha(x,T)}{\partial T }
\frac{\partial n_{Fl}}{\partial E }\right ]_{\Omega,q_x} \nn
 &&\times \int {\frac{d\phi}{2\pi}}
 \frac{v^2 \cos^2 \phi}{ - i\Omega  + i q_x v \cos \phi  + 0^+},
\end{eqnarray}
where $q_x$ is the wavenumber in the Fourier transform of the
current with respect to the $\mb{x}$ coordinate and
$[..]_{\Omega,q_x}$ denotes the $\Omega, q_x$ Fourier component of
the product in the square brackets. We assume that $n_{Fl}$ is
coordinate an time dependent due to the gate electric potential.
At $\Omega w \ll v_F$ the frequency in the denominator of
(\ref{narrow}) can be neglected. Representing $(i q_x v \cos \phi
+ 0^+)^{-1}$ as $\pi \delta (q_x)/(v |\cos\phi|)$ and returning to
the coordinate and time representation we obtain
\begin{equation}\label{narrowj}
{\cal J}_x^y \left(x,T\right) =-\frac{m^*}{2\pi}\int
dx^{\prime}  \frac{\partial \alpha(x^{\prime},T)}{\partial T}
k_{F}(x^{\prime},T),
\end{equation}
where $k_{F}(x^{\prime},T)=\sqrt{2m^{*}(E_F - U(x^{\prime},T))}$.
The above expression is valid in the near vicinity of the gate,
within the length of the electron mean free path. This ballistic
result gives only a part of the spin current, the one associated
with a direct generation action of the time dependent gate. It
does not take into consideration the backflow of diffusion current
due to the spin polarization accumulated on both sides of the
gate. Such a diffusion current and the accumulated polarization
will be calculated within the drift diffusion theory in the next
section. It is interesting to note that besides the AC component,
the spin current also contains the DC component due to time
dependence of $k_F$ in integrand of (\ref{narrowj}). It can be
easily seen that the DC current is obtained from (\ref{narrowj})
if harmonic oscillations of $\alpha(T)$ and $k_F(T)$ are phase
shifted with respect to each other. Such a rectification effect
will be studied in more detail in the next section.


\section{the drift-diffusion equation}

In this section, we are interested in the time dependent spin
dynamics in a disordered system, such that the characteristic
frequency of the gate time dependence is much smaller then the
elastic scattering rate $1/\tau$, while the characteristic length of
the spatial variation of $f_{\b{k}}(\b{R},T)$ is larger than the
mean-free path $l$. To this end, we start from \eq{qkef} and
represent $f_{\b{k}}$ in the form
\begin{equation}\label{Gg}
f_{\b{k}}(\b{R},T) = \mb{\sigma} \cdot \mb{g}_{\b{k}}(\b{R},T).
\end{equation}
Substituting (\ref{Gg}) into (\ref{qkef}) we obtain
\begin{eqnarray}\label{qbg}
&& \frac{\partial \mb{g}_{\b{k}}}{\partial T} + (\b{v} \cdot
\mb{\nabla}_{\bf R}) \mb{g}_{\b{k}} + 2 \alpha_{0} k
(\mb{g}_{\b{k}} \times \mb{h}) + \mb{h} k \frac{\partial
\alpha}{\partial T}\frac{\partial n_{Fl}}{\partial E } \nn &&=
\frac{1}{\tau} \left( \mb{P}_E -\mb{g}_{\b{k}} \right),
\end{eqnarray}
where the unit vector $\mb{h}=(\b{k}\times \b{z})/k$ and $
\mb{P}_E= (2\pi/m^*) \sum_{\b{k}'}\delta(E^{\prime}-E)
\mb{g}_{\b{k}'}$. With this definition of $ \mb{P}_E$, from (\ref{Gg}) and
(\ref{P}), the spin polarization can be expressed as
\begin{equation}\label{PE}
 \mb{P} =\frac{m^*}{2\pi} \int dE \mb{P}_E.
\end{equation}
For simplicity we have assumed small variations of $\alpha$ and put
$\alpha=\alpha_0$ in the third term of the left hand side of
(\ref{qbg}).

In order to derive the drift diffusion equation the function
$\mb{g}_{\b{k}}$ is expressed from \eq{qbg} in the form
\begin{equation}\label{gg}
\mb{g}_{\b{k}} = \hat{\Lambda}^{-1} \left[ \frac{ \mb{P}_E }{\tau}
- \Delta_0 (\mb{g}_{\b{k}}\times \mb{h}) - \mb{h} k \frac{\partial
\alpha}{\partial T} \frac{\partial n_{Fl}}{\partial E } \right],
\end{equation}
where $\hat{\Lambda}$ is the operator inverse to  $ -i\Omega +
\b{v}\cdot\mb{\nabla}_{\bf R} + 1/\tau$. As we mentioned in the
beginning of this section, in the diffusion approximation one can
expand $\hat{\Lambda}^{-1}$ with respect to the small $\Omega $
and $\b{v}\cdot\mb{\nabla}_{\bf R}$ compare to $1/\tau$, so that
\begin{equation}\label{Lambda}
\hat{\Lambda}^{-1} \approx \tau \left[ 1 + i\Omega \tau -
\tau\left(\b{v}\cdot\mb{\nabla}_{\bf R}\right)  +
\tau^2\left(\b{v}\cdot\mb{\nabla}_{\bf R}\right)^2  \right].
\end{equation}
The next step is to express $\mb{g}\times \mb{h}$ in the right hand
side of (\ref{gg}) via $\mb{P}_E$. It can be done by decomposing
$\bm{g}$ into  parallel and perpendicular to $\mb{h}$ parts,
according to
\begin{equation}\label{decompose}
\bm{g}=(\bm{g}\cdot \mb{h})\mb{h} + \bm{g}_{\perp}
\end{equation}
Taking the perpendicular projection of (\ref{qbg}) we find
\begin{equation}\label{perp}
\mb{g}_{\b{k}}\times \mb{h} = \frac{1}{\tau}
(\hat{\Lambda}^{2}+\Delta^{2}_0)^{-1} \left(\Delta_0 \mb{P}_{E\perp}
+ \hat{\Lambda}(\mb{P}_{E}\times\mb{h})\right).
\end{equation}
After inserting this expression into (\ref{gg}) we use the
expansion (\ref{Lambda}) and a similar expansion for
$(\hat{\Lambda}^{2}+\Delta^{2}_0)^{-1}$. It will be further
assumed that $\Delta_0 \ll 1/\tau$ and only the terms not  smaller
than $(\Delta_0 \tau)^2$ will be retained in the right hand side
of \eq{gg}. Since the current source in (\ref{qbg}) is
proportional to $\partial n_{Fl}/\partial E$, for a degenerate
electron gas the function $\mb{P}_{E}$ has a peak at the local
Fermi energy $E_F - U(\b{R}, T)$. Hence, taking into account the
definition (\ref{PE}), this function can be represented as
$\mb{P}_{E}=(2\pi /m^*)\delta(E + U(\b{R}, T) -E_F)\mb{P}$.
Integrating (\eq{gg}) over energies and averaging over the angles
of $\b{k}$ we arrive to the diffusion equation for the spin
transport
\begin{eqnarray}\label{deq}
\frac{\partial \mb{P}}{\partial T} &=& (\bm{\nabla} \cdot D
\bm{\nabla}) \mb{P} - 4Dm^*\alpha_0 \left[\left( \b{z}\times
\bm{\nabla} \right)\times \mb{P}\right]\nn && - \Gamma_s \mb{P} -
\Gamma_s P^z \b{z} -\frac{1}{2}\left(\bm{\nabla}\times \b{z} \right)
\rho \tau \frac{\partial \alpha}{\partial T},
\end{eqnarray}
where $D=v^{2}_{Fl}\tau/2$ is the diffusion constant and
$\Gamma_s=D(m^* \alpha_0)^2$ is the spin relaxation rate. It should
be noted that the diffusion constant can be coordinate and time
dependent via the local Fermi velocity $v_{Fl}$. Except the last
term on the right-hand side of \eq{deq} the above spin diffusion
equation coincides with that derived earlier from the Green function
formalism.~\cite{Malshdiff} In this equation the first three terms
represent, respectively, spin diffusion, spin precession due to SOI
and the D'yakonov-Perel'~\cite{dp} spin relaxation. The last term is
the spin current source provided by the time dependent SOI. The spin
relaxation gives the natural time scale $T_D=\Gamma_D^{-1}$ for the
most of spin diffusion processes. One can also define the
characteristic length of the spin density spatial variations as the
spin diffusion length $L_D=\sqrt{DT_D}$. The drift diffusion
approach is valid when $T_D \gg \tau$. This condition is provided by
the small $\Delta_0$ in comparison with $1/\tau$.

For the following calculations we need an expression for the spin
current. It can be found from its initial definition (\ref{Is}).
Inserting there $f_{\b{k}}$ written in the form (\ref{Gg}), the spin
current is found as
\begin{equation}\label{Jdiff}
{\cal J}_j^i=\sum_{\b{k}}v_j g^i_{\b{k}}(\b{R},T).
\end{equation}
Expressing $g^i_{\b{k}}$ according (\ref{gg}), from
(\ref{Lambda})-(\ref{perp}) we obtain
\begin{eqnarray}\label{JD}
{\cal J}^i_j&=& -D \nabla_j P^i - 2 D m^* \alpha (\delta^{ij} P^z
- P^j\delta^{iz}) \nn && + \frac{1}{2}\varepsilon^{ij3} \rho \tau
\frac{\partial \alpha}{\partial T}.
\end{eqnarray}
This equation shows that the spin current contains three components.
The first and the second terms represent the usual diffusion current
and the current associated with spin precession. These two
contributions have been found in earlier works.~\cite{Malshdiff} The
third term is the new one. It represents the spin current generation
due to the time dependent SOI.

It is interesting to note that in the homogeneous case $\mb{P} = 0$
, and the spin current is simply of the form ${\cal J}^i_j
=\frac{1}{2} \varepsilon^{ij3} \rho \tau (\partial \alpha/\partial
T)$. This result coincides with \eq{sc} by taking the limit of
$\Omega\rightarrow 0$ in the denominator of \eq{sc}.


\subsection{The finite-size AC-biased gate}

We consider an AC-biased gate which is supposed to be infinite
along $\mb{y}$ and with a width $w$ wider then the mean free path
$l$ along $\mb{x}$, with a Rashba coupling constant $\alpha (x)$.
The finiteness of the gate in the $\mb{x}$-direction results in
$\mb{x}$ dependence of $\bm{P}$. At the same time $\bm{P}$ does
not depend on $\mb{y}$. Retaining in \eq{deq} only derivatives with
respect to $\mb{x}$, one can easily see that a pair of equations
for $P^x$ and $P^z$ polarization components is decoupled  from the
equation for $P^y$, and that the current source term enters only
to the latter equation. Therefore, the solution of the diffusion
equation is such that $P^x=P^z=0$ and the equation for the
retained component is
\begin{equation}\label{Py}
\frac{\partial P^y}{\partial T} = \frac{\partial}{\partial x}D
\frac{\partial}{\partial x}P^y - \Gamma_s P^y + \frac{\partial}
{\partial x} S(x,T)
\end{equation}
where the time-dependent spin-current source is given by $S(x,T)
= \rho \tau \partial \alpha/\partial T$.

We will apply \eq{Py} to two problems. In the first problem the
width of the gate will be assumed large, so that $w \gg L_D$. The
opposite case of $w \ll L_D$ will be considered in the second
example.

\subsubsection{$w \gg L_D$}

\begin{figure}[tbp]
\includegraphics[width=.35 \textwidth,angle=0]{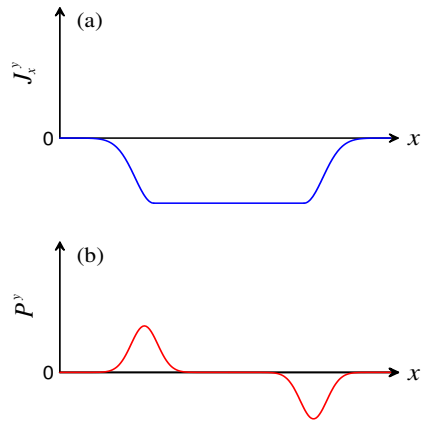}
\caption{Illustration of the spin current flow induced by a wide AC
gate: (a) spin current ${\cal J}_x^y$ as a function of $x$, which is
uniform under the gate and exponentially decays outside the gate;
(b) spin polarization $P^y$ as a function of $x$, which is
accumulated at the gate edges.} \label{WG}
\end{figure}

\begin{figure}[tbp]
\includegraphics[width=.38\textwidth, angle=0]{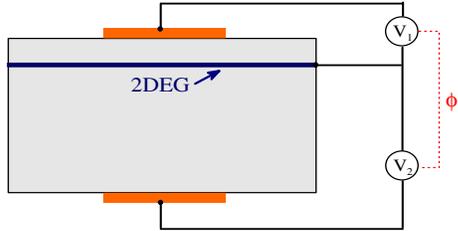}
\caption{Illustration of the setup for spin current generation. The
front gate is close to 2DEG. This gate controls the spin-orbit
coupling constant via periodic oscillations of $V_1$. $\phi$ denotes
the phase shift between electric potentials of the front and back
gates.} \label{DCSC}
\end{figure}

Let us assume that in the interval $x_1 <x < x_2$ the diffusion
constant is $D(T)$ and $\alpha = \alpha_0 + \alpha(T)$, while
outside this interval $D=D_0$ and  $\alpha = \alpha_0 $. We will
consider relatively slow variations in time of $\alpha(T)$, such
that their frequency is much less than $\Gamma_s$. Consequently, the
time derivative on the left-hand side of \eq{Py} will be neglected.
The spin current induced in the range $x_1 <x < x_2$ will be
injected near edges at the points $x_1$ and $x_2$, and the injected
polarization will diffuse outward, as well as backward to the
modulation gate region, as shown in Fig.~1. For analysis of this
injection process it is enough to consider a vicinity of either
edges. Let it be the right edge. Then, the solution of \eq{Py} has
the form
\begin{equation}\label{sol1}
P^y = A \exp\left(-\frac{|x-x_2|}{L_D}\right),
\end{equation}
where $L_D=(\alpha_0 m^*)^{-1}$. The coefficient $A$, in its turn,
can be found from the spin current conservation at the boundary
$x_2$. On the right of the boundary the current is represented by
the diffusion current $AD_0/L_D$. It must be equal to the current
on the left, which is the sum of the diffusion and the source
terms $-AD(T)/L_D-S(T)$. The factor $A$ is thus
\begin{equation}\label{A}
A=-\frac{\rho(T) \tau }{2m^* \alpha_0\left(D_0 +
D(T)\right)}\frac{\partial \alpha}{ \partial T}.
\end{equation}
Using the expression $D=v_F^2\tau/2$ and $\rho=k_F^2/2\pi$ we
obtain the polarization at the right edge
\begin{equation}\label{P1}
P^y=-\frac{m^*\rho(T)}{\pi \hbar \alpha_0 (\rho(T)+\rho_0)}\frac{\partial
\alpha}{\partial T}.
\end{equation}
We restored a conventional dimensionality by writing $\hbar$ in a
proper place. An important feature of this expression is that $P^y$
does not depend on the absolute value of the SOI coupling constant,
but rather from the relative amplitude of its variations in time.
Hence, the same spin injection effect is expected for both,  QW with
the large Rashba SOI, and QW with not large SOI, providing that the
relative variations of $\alpha$ in time are the same in both cases.
Another interesting phenomenon, which follows from (\ref{P1}), is
that the AC modulated SOI can result in stationary spin accumulation
near the gate edges. This is due to time dependence of the electron
density in \eq{P1}. The DC spin density is obtained by time
averaging of this expression. A phase shift between $\rho(T)$ and
$\alpha(T)$ is required to have this average nonzero. Such a shift
can be achieved, for example, by manipulating the front and back
gates, as shown in \fig{DCSC}.  Both the back and the front gates
are equally efficient to induce electron density oscillations. At
the same time, as shown by Grundler,~\cite{Grun00} the front gate
close to 2DEG is necessary to control efficiently $\alpha$.
Therefore, the required phase shift can be obtained by choosing
appropriate phases of $V_1$ and $V_2$.

\subsubsection{$w \ll L_D$}

Another interesting regime is the case of narrow gate with size $w$
(in $x$ direction) much shorter then the characteristic diffusion
length $L_D$. Outside the gate region the solution of \eq{Py} has
the same form as in the previous example, namely,
\begin{equation}\label{sol2}
P^y = \pm A \exp\left(- \frac{|x|}{L_D}\right),
\end{equation}
where the - and + signs refer to $x>0$ and $x<0$ regions,
respectively. Inside the gated region, as seen from Fig.~2 near
$x=0$, the polarization  varies very fast. Therefore, within this
range one can retain in \eq{Py} only terms containing derivatives
of $P^y$ and $S$. This gives
\begin{equation}\label{conserv}
-D(x)\frac{\partial P^y}{\partial x} - S(x)= C.
\end{equation}
This equation has a form of the spin current conservation low. The
constant $C$ is, obviously, equal to the current just outside the
gated region, where $S(x)=0$, but still $x \ll L_D$. Hence, from
\eq{sol2} one obtains $C=D_0A/L_D$. Further, integrating
\eq{conserv} we arrive to the expression for $P^y$ in the near
vicinity of $x=0$
\begin{eqnarray}\label{Px}
P^y&=&-\frac{A}{L_D}x  - \int_0^x dx' \frac{S(x')}{D(x')} -
\frac{A}{L_D}\int_0^x dx' \nn && \times
\left(\frac{D_0}{D(x')}-1\right).
\end{eqnarray}
This expression must coincide with \eq{sol2} at $w<x \ll L_D$,
namely, with $A\exp(-x/L_D)\simeq A-Ax/L_D$. Substituting the latter
instead of $P^y$ into \eq{Px} and setting $x \rightarrow \infty$,
one obtains
\begin{eqnarray}\label{A1}
A&=&-\left[1+\frac{1}{L_D}\int_0^{\infty} dx
\left(\frac{D_0}{D(x)}-1\right)\right]^{-1} \nn && \times
\int_0^{\infty} dx \frac{S(x)}{D(x)}\, .
\end{eqnarray}

One can neglect the integral in the square brackets because it is
small by the parameter $w/L_D$, except of special cases when
$D_0/D(x)$ becomes very large in some points. Expressing $D(x)$ via $\rho(x)$,
as it has been done in the previous
problem, we arrive to the simple formula
\begin{equation}\label{Afin}
A= - \frac{m^{*2}}{\pi\hbar^3}\int dx \frac{\partial
\alpha}{\partial T}.
\end{equation}
Since the integral in \eq{Afin} is of the order of $w d\alpha/dT $,
the polarization injected in the case of a narrow gate is smaller by
a parameter $w/L_D$ than in the previous case of the wide gate. That
is because of a strong counterflow of the diffusion current reducing
the effect of the current $-S$ pumped by the gate. Also, unlike the
previous example, the narrow gate can not inject the stationary spin
polarization. The rectification effect is absent in this case
because in \eq{Afin} there are no time dependent parameters beside
$\partial \alpha/\partial T$.

\begin{figure}[tbp]
\includegraphics[width=.35 \textwidth,angle=0]{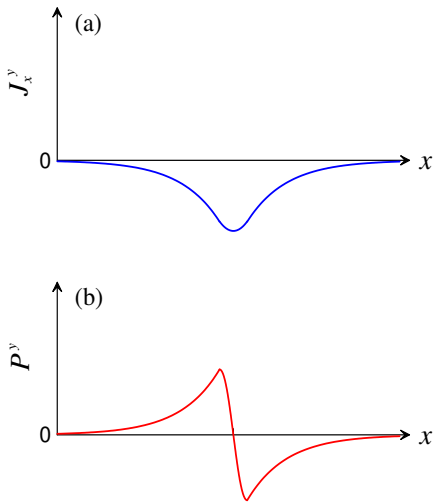}
\caption{Illustration of the spin current flow induced by a narrow
AC gate: (a) spin current ${\cal J}_x^y$ as a function of $x$, which
is exponentially decays outside the gate; (b) spin polarization
$P^y$ as a function of $x$, which is accumulated around the
AC-biased narrow gate. Polarization has opposite signs on two sides
of the gate.} \label{NG}
\end{figure}

Figure~\ref{NG} illustrates the spin current [\fig{NG}(a)] and spin
polarization [\fig{NG}(b)] induced by the narrow AC gate.  The
diffusion current exponentially decays far from the gate region with
spin accumulation shown in \fig{NG}(b). The spin polarization has
opposite signs on two sides of the gate.

\subsection{Spin flow in a channel}

We consider a 2D channel (strip) of the width $d$, so that particles
are confined in the $y$ direction and free to move in the
$x$-direction. The time dependent gate atop of this channel causes
time variations of the Rashba SOI. We assume $d \gg l$ and apply the
drift-diffusion equation to describe the spin transport in the
channel. Since the gate is supposed to be infinite in the $x$
direction, the spin polarization does not depend on $x$ and only
derivatives with respect to $y$ and $T$ have to be retained in
\eq{deq}.

In this case one can look for a solution of \eq{P1} such that $P^y =
P^z = 0$ and $P^x = P^x(y)$. It is easily to check that with such
choice of $\mb{P}$ the spin current \eq{JD} in the $y$ direction
does not contain $y$ and $z$ polarization components. The
homogeneous $y$-polarized current given by the last term in \eq{deq}
flows only in the $x$ direction. Hence, the only current flowing
along the $y$ axis is $x$-polarized. For this current the boundary
condition is imposed that $({\cal J}^{x}_{y})_{x=\pm d/2}=0$. Since
the time dependent $\alpha$ in the last term of \eq{JD} generates
the spin current ${\cal J}^{x}_{y}=(1/2) \rho\tau
\partial \alpha/\partial T$, to satisfy the boundary condition
this current must can be compensated by counterflow of the spin
diffusion current given by the first term in \eq{JD}. Hence, the
boundary condition is
\begin{equation}\label{BC}
\left[D \nabla_y P^x - \frac{1}{2}\rho \tau \frac{\partial
\alpha}{\partial T}\right]_{x=\pm \frac{d}{2}}=0.
\end{equation}
In its turn, the equation for $P^x$ has the form
\begin{equation}\label{PxF}
\frac{\partial P^x}{\partial T} = D(T) \frac{\partial^2
P^x}{\partial y^2} - \Gamma_s P^x (y).
\end{equation}

Assuming that the frequency of $\alpha$ variations is small
compare to the spin relaxation rate $\Gamma_s$, we obtain the
general solution of \eq{PxF} in the form
\begin{equation} \label{PxCC}
P^x (y,\Omega) = C_1 e^{\kappa y} + C_2 e^{-\kappa y},
\end{equation}
where $\kappa = 1/L_D$. The solution satisfying the boundary
condition is easily obtained as
\begin{equation}
P^x(y,T) = \frac{m^*}{2\hbar \pi \alpha}  \frac{\sinh \left(\kappa
y\right)}{\cosh \left( \kappa \frac{d}{2} \right)} \frac{\partial
\alpha}{\partial T}.
\end{equation}
For an extended 2D electron system with $d\rightarrow \infty$, it is
easy to see that $P^x(y,T)\rightarrow 0$ indicating the absence of
the bulk spin density, in agreement with the result of section IIIA.
At the same time, as expected, the polarization is accumulated near
$y=\pm d/2$, decreasing exponentially when the distance from the
boundary increases.

\section{Summary and discussion}
\label{summary}

We have employed the electric gate effect on the Rashba spin-orbit
interaction in narrow gap semiconductor QW and considered spin
transport in a 2DEG with the space-time dependent Rashba spin-orbit
interaction. The variations of SOI in time and space were assumed to
be provided by gates of various shapes. Spin transport was
considered in the framework of the Keldysh formalism which has been
applied to derive the Boltzmann equation for the spin distribution
function in the phase space. This equation was further employed in
derivation of the drift diffusion equation for the spin density. We
found out that besides the usual terms, both the Boltzmann and drift
diffusion equations contain the spin current motive force
proportional to the time derivative of the SO coupling parameter.

We have considered several examples with various gate geometries.
Although these examples do not embrace many other interesting
possibilities, nevertheless, they demonstrate how spin current flow
and spin density accumulation can be electrically controlled by
means of gates. It has been shown that in some geometries the AC
bias applied to the gate can result in the DC spin current and the
stationary spin accumulation. This is a simplest self-rectification
effect. Probably, more efficient could be a special rectification
setup consisting of several gates in series and combinations of back
and forward gates to control separately electron density and SO
parameter.

\acknowledgments{This work was partly funded by the Taiwan National
Science Council (NSC) under grant Nos. NSC93-2119-M-007-002 (NCTS);
the Russian Academy of Sciences and the RFBR grant No. 03-02-17452;
and the Swedish Royal Academy of Science. A.G.M. acknowledges the
hospitality of NCTS where this work has been initiated.

\end{document}